**Mechanisms and timing of carbonaceous chondrite delivery to the Earth**

Francis Nimmo[1*], Thorsten Kleine[2], Alessandro Morbidelli[3], David Nesvorny[4]

[1]Dept. Earth & Planetary Sciences, University of California Santa Cruz, Santa Cruz CA 95064, USA

[2]Max Planck Institute for Solar System Research, 37077 Gottingen, Germany

[3]College de France, 75 231 Paris Cedex 05, France

[4]Dept. Space Studies, Southwest Research Institute, Boulder CO 80302, USA

*Corresponding author: fnimmo@ucsc.edu

**Abstract**

The nucleosynthetic isotope signatures of meteorites and the bulk silicate Earth (BSE) indicate that Earth consists of a mixture of "carbonaceous" (CC) and "non-carbonaceous" (NC) materials. We show that the fraction of CC material recorded in the isotopic composition of the BSE varies for different elements, and depends on the element's tendency to partition into metal and its volatility. The observed behavior indicates that the majority of material accreted to the Earth was NC-dominated, but that CC-dominated material enriched in moderately volatile elements by a factor of ~10 was delivered during the last ~2-10% of Earth's accretion. The late delivery of CC material to Earth contrasts with dynamical evidence for the early implantation of CC objects into the inner solar system during the growth and migration of the giant planets. This, together with the NC-dominated nature of both Earth's late veneer and bulk Mars, suggests that material scattered inwards had the bulk of its mass concentrated in a few, large CC embryos rather than in smaller planetesimals. We propose that Earth accreted a few of these CC embryos while Mars did not, and that at least one of the CC embryos impacted Earth relatively late (when accretion was 90-98% complete). This scenario is consistent with the subsequent Moon-forming impact of a large NC body, as long as this impact did not re-homogenize the entire Earth's mantle.

1. Introduction

A key question regarding the growth of the Earth is the original location (provenance) of the materials it grew from, and when they were delivered. Answering this question would allow us to differentiate between different accretion scenarios, and to understand the origin of terrestrial water (and other volatile species). One way of investigating this problem is to use the fact that Solar System materials are divided into two different groups having distinct nucleosynthetic isotopic signatures (Budde et al., 2016; Warren, 2011): non-carbonaceous ("NC") meteorites (e.g., angrites, eucrites, ureilites, several iron meteorite groups, main-group pallasites, enstatite and ordinary chondrites), presumed to originate in the inner solar system, and carbonaceous ("CC") meteorites (e.g., carbonaceous chondrites, some other groups of irons, Eagle Station pallasites), presumed to originate from beyond the asteroid belt.

Nucleosynthetic isotope anomalies can be used to constrain the provenance of Earth's accreted material either by tying its building material to specific groups of meteorites (e.g., Dauphas et al., 2014; Warren, 2011) or, more generally, by quantifying the relative contribution of NC and CC materials to Earth (e.g., Burkhardt et al., 2021; Dauphas, 2017; Schiller et al., 2018; Warren, 2011). While the isotopic signatures of individual elements may be interpreted to reflect a large fraction of CC material in Earth (a few tens of percent) (Schiller et al., 2020, 2018; Warren, 2011), the isotope signatures of a comprehensive set of elements combined indicates that the CC fraction in Earth is in fact quite low, on the order of a few percent (Burkhardt et al., 2021; Dauphas, 2017; Dauphas et al., 2024; Martins et al., 2023; Render et al., 2022; Savage et al.,

2022; Steller et al., 2022). Moreover, the isotopic composition of the bulk silicate Earth (BSE) most closely resembles that of enstatite chondrites, suggesting that Earth predominantly accreted from material having an enstatite chondrite-like isotopic composition, either because Earth grew homogeneously from such material (Dauphas, 2017; Dauphas et al., 2024, 2014) or because enstatite chondrites happen to reflect the average isotopic composition of Earth's isotopically heterogeneous building materials (Burkhardt et al., 2021; Steller et al., 2022). It is also clear, however, that Earth incorporated material that remained unsampled among meteorites (Budde et al., 2019; Burkhardt et al., 2021, 2016, 2011; Fischer-Gödde et al., 2020; Render et al., 2022, 2017) and which probably derives from the innermost disk (Burkhardt et al., 2021). Thus, the isotopic composition of Earth cannot be reconstructed solely as a combination of known meteorites.

Dauphas (2017) showed that for siderophile elements, the BSE's isotopic composition predominantly reflects the provenance of Earth's late-stage building blocks. This observation has been used to reconstruct how the provenance of accreted materials has changed over time and, in particular, when CC material was accreted by Earth. For instance, by assuming that Earth can be built from known chondrites, Dauphas (2017) argued that Earth predominantly accreted from material with an enstatite chondrite-like isotopic composition, and that a small fraction of carbonaceous chondrite-like material was added only during the first ~60% of accretion. Using a larger and more comprehensive set of elements, Dauphas et al. (2024) revised this conclusion and argued that the first ~60% of Earth's accretion consisted virtually solely of material with an enstatite chondrite-like isotopic composition, while a small fraction of CI chondrite-like material was added only during the last ~40% of accretion. By solely considering Mo, a moderately siderophile element that records only the last ~10% of Earth's accretion, Budde et al. (2019) found a large fraction of ~46% CC material for this growth stage, but argued that the total CC fraction in the bulk Earth is much lower. Using Fe isotopes, Schiller et al. (2020) argued that all the BSE's Fe derives from CI chondrite-like material, implying the late-stage addition of a relatively large fraction of pure CC material in Earth. This conclusion, however, is inconsistent with the aforementioned mixed NC-CC heritage of the BSE's Mo (Budde et al., 2019), and with the BSE's Ni isotopic composition which is distinct from CI chondrites (Hopp et al., 2022b). Thus, while most studies agree that the overall CC contribution to Earth is rather low, there is considerable uncertainty about how and when precisely this CC material was added to the growing Earth.

Despite these uncertainties, it is clear that the Earth itself appears to be a mixture of NC and CC materials, and that this mixing ratio in the BSE varies from element to element. In this paper we will use the observation that the CC fraction recorded is larger for more siderophile elements to infer how and when the Earth accreted CC material. Recently, Dauphas et al. (2024) carried out an analysis that is similar in some ways to ours, but also differs in two key aspects. First, while Dauphas et al. (2024) fix the boundaries between different accretion stages *a priori*, we allow them to vary. As we will show, this is essential to more precisely define when and how the Earth accreted CC material. Second, we also treat Zn, which is moderately volatile, as distinct from the refractory elements. As CC material is likely volatile-rich relative to NC material, the contrasting behaviour observed for refractory compared with volatile elements tells us how and when Earth acquired its volatile elements.

## 2. Observations

In multi-element isotope space, meteorites plot in two distinct fields defining the NC-CC dichotomy (e.g., Kleine et al., 2020; Kruijer et al., 2020b; Warren, 2011). In addition, there are isotopic variations within each reservoir, which for the NC reservoir are linearly correlated with ureilites and some iron meteorites defining one end of the NC trend, and enstatite chondrites and Earth the other (Burkhardt et al., 2021; Spitzer et al., 2020) (Fig. 1). The isotopic variations among NC and CC meteorites thus indicate the presence of at least three distinct components that may have contributed to the building material of the terrestrial planets, namely the two endmembers defining the NC trend, and CC meteorites. For some elements (e.g., Mo, Zr), Earth plots at one end of the NC trend as defined by meteorites (Budde et al., 2019; Burkhardt et al., 2021; Render et al., 2022), indicating that one of these components remains unidentified among known meteorites (Fig. 1).

An important recent discovery is the finding of nucleosynthetic isotope anomalies for the moderately volatile element Zn (Martins et al., 2023; Paquet et al., 2022; Savage et al., 2022; Steller et al., 2022). For Zn, the BSE has a mixed NC-CC composition, indicating that ~70% of terrestrial

Zn derives from the NC, and ~30% from the CC reservoir (Savage et al., 2022; Steller et al., 2022). Martins et al. (2023) inferred a somewhat larger fraction of ~50% CC-derived Zn. Importantly, in a plot of $\mu^{66}$Zn versus $\mu^{54}$Cr, the BSE plots off the NC trend towards the CC field (here µ refers to the ppm deviation from terrestrial standard ratios). Figure 1 shows a mixing line connecting the NC and CC fields that pass through the BSE composition, where the curvature of this mixing line depends on the Zn/Cr ratios of the NC and CC endmembers. The mixing line shown is strongly curved because of the volatile-rich nature of CI chondrites and the overall volatile element-depleted nature of the BSE. Such mixing lines would become straight lines only if two reservoirs had identical Zn/Cr ratios. The latter option can be excluded, however, because there is no reason why the NC and CC material accreted by Earth should all have had the same degree of volatile element depletion, and one would expect that the CC material was volatile-rich compared to the NC material, given that it formed at greater heliocentric distance.

An important observation from the mixing line shown in Fig. 1 is that it originates close to the composition of enstatite chondrites on the NC trend. This observation holds regardless of which carbonaceous chondrite group is used for the CC endmember and, as such, provides strong evidence that the *average* composition of the NC material accreted by Earth has approximately the isotopic composition of enstatite chondrites (Steller et al., 2022). This allows us to treat the building material of Earth as a two-component mixture: (1) NC material with an on average enstatite chondrite-like isotopic composition; and (2) CC material. In reality, Earth almost certainly accreted from isotopically more heterogeneous materials. However, since all the isotopic variations in the NC reservoir are linearly correlated, what the results are sensitive to is the averaged isotopic composition, which is approximately the same as that of enstatite chondrites. This allows us to treat the NC material in Earth as a single component and affords a significant simplification compared with Dauphas et al. (2024).

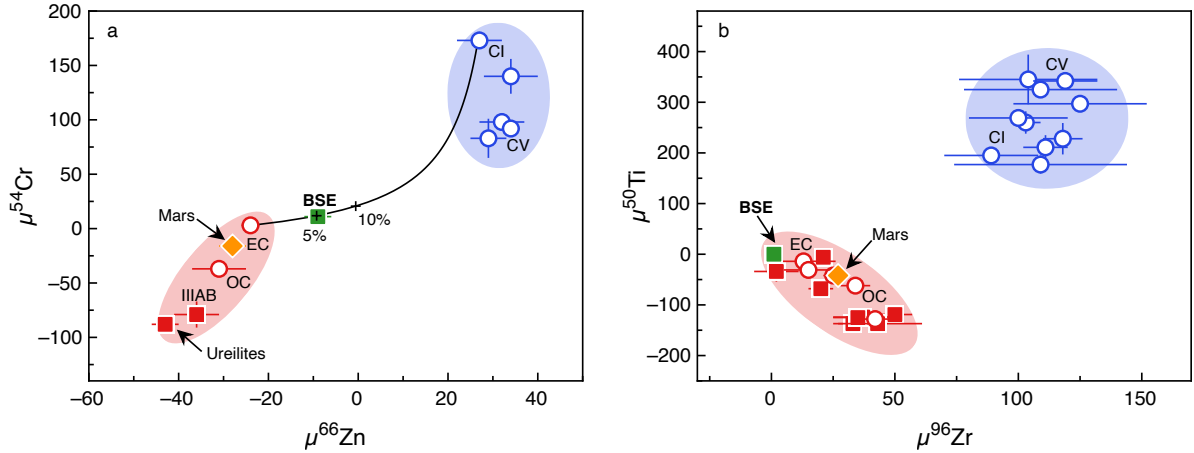

**Fig 1.** $\mu^{66}Zn$ versus $\mu^{54}Cr$ (a) and $\mu^{96}Zr$ versus $\mu^{50}Ti$ (b) for meteorites, the BSE, and Mars. Both plots show the dichotomy between NC and CC meteorites and the correlated nature of isotope anomalies in the NC reservoir. a) Owing to the mixed NC-CC heritage of terrestrial Zn, the BSE plots off the NC trend towards the CC field. A mixing line between the BSE and CI chondrites intersects the NC trend at the composition of enstatite chondrites (EC), indicating that the average NC material in Earth has an EC-like isotopic composition. The mixing line also shows that Earth contains ~5-6% CI chondrite-like material. The NC concentration of Zn (38 µg/g) is calculated from the Zn concentrations in the BSE and CI chondrites [54 and 309 µg/g, respectively; Palme and O'Neill (2014)]. Mars plots within the NC field, indicating a smaller CC contribution compared with the Earth. b) Earth plots at the upper end of the NC trend indicating it incorporated material with lower $\mu^{96}Zr$ (i.e., *s*-process-enriched material) and higher $\mu^{50}Ti$; this material remained unsampled among meteorites. Plots slightly modified from Kleine and Nimmo (2024). Data sources: Cr, Ti [meteorites from the compilation of Spitzer et al. (2020); Mars from Kruijer et al. (2020a) and Burkhardt et al. (2021)]; Zr (Render et al., 2022); Zn (Kleine et al., 2023; Steller et al., 2022).

Although the BSE's isotopic composition is most similar to that of enstatite chondrites, it is not exactly the same. These small deviations (typically on the order of <15 ppm) cannot reflect variations along the NC trend, because the BSE plots off this trend in the $\mu^{66}Zn$–$\mu^{54}Cr$ plot (Fig. 1). The fraction of CC-derived element in the BSE (not be confused with the bulk CC fraction in Earth) can thus be calculated by mass balance from the isotopic difference between the BSE and enstatite chondrites as follows:

$$f_{CC} = \frac{\mu^i E_{BSE} - \mu^i E_{EC}}{\mu^i E_{CC} - \mu^i E_{EC}} \qquad (1)$$

where $\mu^i E$ refers to the isotope anomaly for element $E$ in the BSE, enstatite chondrites (EC), or carbonaceous chondrites (CC). For the CC component, we assume the isotopic composition of CI chondrites, for the following reasons. First, using this composition we calculate CC-derived fractions of Ti and Cr in the BSE of 0.07±0.07 and 0.03±0.04, respectively (Table 1). These two elements can be considered lithophile for most of Earth's accretion, and so the CC fractions of these elements provide a good estimate for the bulk CC fraction in Earth of ~0.05. This estimate is in excellent agreement with a fraction of 0.06±0.02 CI material in Earth estimated from the Zn isotopic compositions and concentrations of the BSE and CI chondrites (Savage et al., 2022; Steller et al., 2022), and also with a CC mass fraction in Earth of ~0.04 estimated from the position of the BSE in multi-element isotope plots (Burkhardt et al., 2021). If instead we use CV chondrites to represent the CC material in Earth, the Zn isotope data would indicate a CC mass fraction of 0.15±0.06, while Ti and Cr isotopes yield CC fractions of 0.05±0.04 and 0.06±0.10, respectively. Thus, unlike for CI chondrites, using CV chondrites as a proxy for the CC component results in inconsistent CC fractions for Zn compared to Ti and Cr. Thus, using CI chondrites to represent the CC material accreted by Earth provides the most consistent results.

Our approach for calculating CC fractions can be applied to several elements, including Ti, Cr, Ni, Zn, and Mo (Table 1). However, for Fe, Si, Zr, W, and Ru this calculation is not possible. For Fe, the BSE's isotopic composition is indistinguishable from that of CI chondrites, and the difference to enstatite chondrites is quite small (Hopp et al., 2022a, 2022b; Schiller et al., 2020). This results in a very large uncertainty on the calculated CC fraction, which makes Fe unsuitable for our approach. For Si isotopes any difference between enstatite chondrites and the BSE depends on how mass-dependent Si isotope variations are corrected for (Dauphas et al., 2024; Onyett et al., 2023). For Zr and Ru the BSE's isotopic composition deviates from enstatite chondrites (Fischer-Gödde and Kleine, 2017; Render et al., 2022), but this deviation at least in part reflects the enrichment of Earth in *s*-process nuclides compared to meteorites, and not solely the addition of CC material. For Mo this is also the case, but this problem can be overcome by using $\Delta^{95}$Mo values (see Supplementary Materials). A similar approach cannot be taken for Zr, which has an insufficient number of isotopes, while for Ru sufficiently high-precision measurements are not yet available (Bermingham et al., 2018; Fischer-Gödde and Kleine, 2017). Similar problems exist for W, where given the *s*-process-enriched nature of the BSE, meteorites should exhibit $^{183}$W excesses. Such excesses, however, have so far only been identified for CC, but not for NC meteorites (e.g., Chiappe et al., 2023; Kruijer et al., 2017; Qin et al., 2008). In addition, no $^{183}$W difference between NC meteorites and the BSE has yet been resolved. This may reflect a mixed NC-CC heritage of the BSE's W, where the expected $^{183}$W deficit in the BSE compared to NC meteorites is counterbalanced by the addition of CC-derived W having an $^{183}$W excess. However, given the lack of resolved $^{183}$W anomalies among NC meteorites, this is difficult to quantify, making W isotopes unsuitable for our approach of calculating CC fractions.

**Table 1:** Isotope compositions the BSE and enstatite and CI chondrites used to calculate CC fractions.

| Isotope anomaly[a] | CI chondrites | Enstatite chondrites | BSE | $f_{CC}$ | $D_{core/mantle}$[b] |
|---|---|---|---|---|---|
| $\mu^{50}$Ti | 185±20 | -17±6 | -2±12 | 0.07±0.07 | 0 |
| $\mu^{54}$Cr | 165±17 | 3±5 | 8±7 | 0.03±0.04 | 3 |
| $\mu^{62}$Ni | 23±4 | 0±3 | 3±2 | 0.13±0.15 | 26 |
| $\Delta^{95}$Mo[c] | 26±2 | -7±4 | 7±5 | 0.42±0.17 | 174 |
| $\mu^{66}$Zn | 27±5 | -24±3 | -9±2 | 0.29±0.07 | 0 |

[a]Data sources for the isotope anomalies are given in the Supplementary Information.
[b]$D$ is the single stage metal-silicate partition coefficient calculated from the present-day BSE abundances and the estimated bulk Earth composition (Palme and O'Neill, 2014). See Supplement for details.
[c]$\Delta^{95}$Mo is defined as the deviation from an *s*-process mixing line in the $\mu^{94}$Mo-$\mu^{95}$Mo plot passing through the origin: $\Delta^{95}$Mo = $\mu^{95}$Mo – 0.596 × $\mu^{94}$Mo (Budde et al., 2019). For CI chondrites we use the characteristic $\Delta^{95}$Mo value of the CC reservoir reported in Budde et al. (2019).

Not including Fe and Zr in our model is unproblematic because these elements, given their geochemical and cosmochemical behavior, are expected to provide the same information as Ni and Ti, respectively. Including W would have been useful, as its siderophility is intermediate between Ni and Mo, and so the CC fraction recorded in W should also be between those of Ni and Mo. As noted above, the BSE's W likely has a mixed NC-CC heritage, but this is difficult to say with certainty. More problematic is that Ru cannot easily be included in our model, because Ru isotopes are uniquely useful for providing information on the isotopic nature of the late veneer, the final ~0.5% mass delivered to Earth (e.g., Chou, 1978). Nevertheless, while the CC fraction cannot be calculated from the Ru isotopic data using equation (1), it has been inferred indirectly using Ru and Mo isotopic data for lunar impactites. These formed during basin-forming impacts on the

Moon and, given the low siderophile element content of the lunar crust, their Ru and Mo isotopic composition reflect those of the impactors. As shown by Worsham and Kleine (2021), lunar impactites analyzed to date have the same average Ru isotopic composition as the BSE, indicating that basin-forming impactors on the Moon, although representing a much smaller mass compared to the entire late accretionary additions to Earth, come from the same population of bodies as the late veneer. As such, the Mo isotopic composition of the lunar impactites may provide a good approximation of the average isotopic composition of the late veneer. Importantly, the Mo isotopic composition of the impactites overlaps with those of enstatite chondrites, suggesting that any contribution of CC material to the late veneer was smaller than 20% (Worsham and Kleine, 2021).

The CC contributions for each element calculated using equ. (1) reveal some systematic differences. For non-volatile lithophile elements (Cr, Ti), the CC fractions are low, on the order of a few percent, but for the moderately volatile lithophile element Zn the CC fractions are substantially larger (Table 1). For the non-volatile siderophile elements Ni and Mo the CC fractions are also larger and increase with an element's siderophility from Ni to Mo (although we note that the CC fraction for Ni is quite uncertain, owing to the only small overall $\mu^{62}$Ni variations, Table 1). As we will show below, these systematic differences indicate that Earth as a whole received only small contributions from CC material and that this CC material was accreted primarily during the late stages of accretion, and delivered a significant share of Earth's moderately volatile element budget.

## 3. Modeling Approach

To determine the fraction of CC material in the BSE (equivalently, "mantle") as recorded by elements of different siderophility we will use the approach developed by Dauphas (2017). In the Dauphas formulation, the fraction of atoms at present in the mantle that were delivered when the planet grew from fraction $x_1$ to $x_2$ of its present mass is given by

$$\int_{x_1}^{x_2} (1 + \kappa) x^\kappa \, dx \qquad (2)$$

where $x$ is the instantaneous mass fraction (in units of the final planet mass) and $\kappa$ is given by

$$\kappa = \frac{D\gamma k}{1-\gamma} \quad (3)$$

where $D$ is the metal-silicate partition coefficient of the element in question, $\gamma$ is the core mass fraction (assumed 0.33 for all bodies) and $k$ is the effective fraction of each impactor core that equilibrates with the proto-planet mantle (Deguen et al., 2014). The entire mantle is assumed to take part in this equilibration process. Note that this formulation assumes that $D$, $\gamma$, and $k$ are all constant, and that accretion occurs by a succession of small impacts; relaxation of these assumptions does not greatly change the results (Dauphas, 2017).

To track the fraction of CC-derived atoms in the present-day mantle, $f_{CC}$, using equation (2) gives

$$f_{cc} = \int_0^1 f(x)(1+\kappa)x^{\kappa}\,dx \quad (4)$$

where $f(x)$ is the instantaneous fraction of CC material delivered when the protoplanet is at a mass fraction $x$.

Below we will generally assume a simple three-step form for $f(x)$, with the three epochs referred to as the main stage of accretion, late-stage accretion, and the late veneer. This is undoubtedly a simplification of the real picture, but there is some dynamical support for abrupt changes in the provenance of material delivered. For instance, both O'Brien et al. (2006) and Raymond et al. (2006) found that the delivery of bodies with more distant initial positions to the terrestrial planet-forming zone typically happened late in the accretion sequence. Similarly, O'Brien et al. (2014) found that in "Grand Tack" models, the fraction of outer solar system material accreted to the terrestrial planets increased markedly towards the end of accretion. Assuming that CC material originates from a distant reservoir, these results suggest that a late increase in the CC fraction delivered is not an unreasonable assumption.

For our modeling approach we assume that the instantaneous CC fraction is held constant at $f_0$ up to some characteristic mass $x_0$ (main-stage accretion), and then held constant at $f_1$ in the mass range $x_0$ to $x_1$ (late-stage accretion). Above $x_1$ the CC fraction is set to $f_2$, representing the late veneer (Fig. 2). A non-zero value of $f_0$ would imply that the early stages of accretion included some CC material. Our approach differs from that of Dauphas et al. (2024) in that we allow the location of the boundary $x_0$ to vary, rather than holding it fixed *a priori*. We will also initially assume that the element concentrations delivered by bodies from the two reservoirs are the same; we relax this assumption when discussing volatile elements like Zn (Section 4.2).

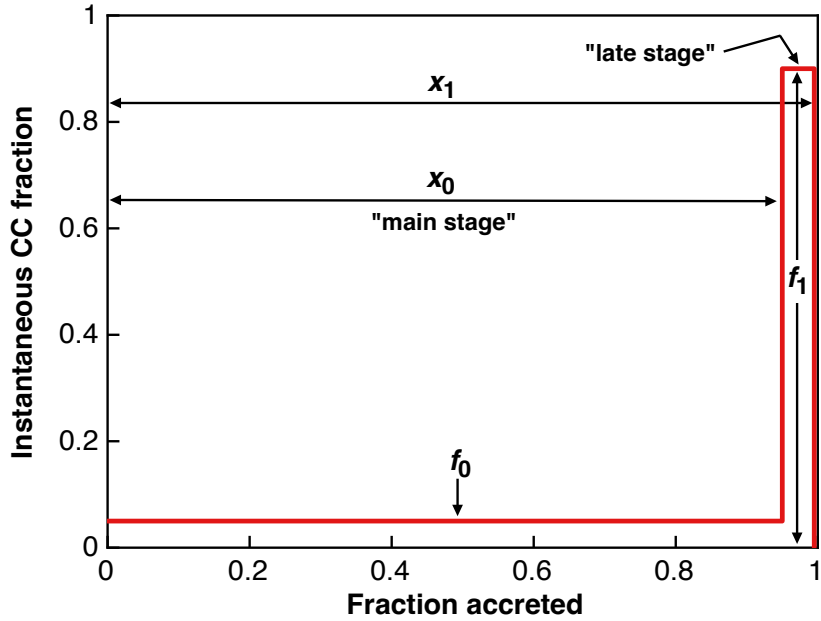

**Figure 2.** Sketch of accretion model. The Earth accretes material with a CC fraction of $f_0$ until it has reached $x_0$ of its final mass ("main stage"); thereafter it accretes material with a CC fraction of $f_1$ up to $x_1$ of its final mass ("late stage"). The final stage (the late veneer) starts at mass $x_1$ and has a CC fraction of $f_2$. In this work we set $x_1$=0.995 and $f_2$=0.

For our model, integration of equation (4) yields

$$f_{cc} = f_2 + x_1^{\kappa+1}(f_1 - f_2) + x_0^{\kappa+1}(f_0 - f_1) \qquad (5)$$

This expression has the correct limiting form in the case that $x_1$=1 (no late veneer). Less obviously, if an element is very siderophile, then $\kappa$>>1 in which case $f_{cc}=f_2$, which correctly indicates that for highly siderophile elements, the BSE will record the CC fraction for only the very last material added (Dauphas, 2017). Conversely, if the element is lithophile then $\kappa$=0 and equation 5 yields $f_2+x_1(f_1-f_2)+x_0(f_0-f_1)$, which is simply a weighted average of all the material going into the mantle. In all our models below we assume that the late veneer is the last 0.5% of the Earth's mass ($x_1$=0.995), based on siderophile element concentrations (e.g., Walker, 2009), and that it is pure NC ($f_2$=0), based on Ru isotopes (Worsham and Kleine, 2021).

For lithophile elements and neglecting the role of the late veneer, a more general expression for the BSE's CC fraction $f_{CC}$ is given by

$$f_{CC} = \frac{x_0 f_0 + v_{fac} f_1 (1-x_0)}{x_0 + v_{fac}(1-x_0)} \quad (6)$$

where $v_{fac}=C_1/C_0$ is the ratio of the elemental concentration in the late-added material $C_1$ compared to that in the early material $C_0$. For refractory elements we expect $v_{fac}=1$, while for volatile elements $v_{fac}$ may be different from 1. Similarly, the final BSE elemental concentration $C'$ is given by

$$C' = C_0(x_0 + v_{fac}(1-x_0)) \quad (7)$$

Equation (5) can be readily generalized to a multi-step case, which is useful for analyzing the output of N-body accretion codes. In this case, we have

$$f_{cc} = f_N + \sum_{i=0}^{N-1} x_i^{\kappa+1}(f_i - f_{i+1}) \quad (8)$$

Based on Ti and Cr isotopes the weighted average CC fraction $f_{CC}$ is approximately 6±4% for Earth (Table 1). Similarly, based on Mo isotopes we know that $f_1$ is 25% or more. If the main phase of accretion is purely NC, then $f_0=0$.

Given an assumed $f_0$ and $f_1$, we can search for combinations of $x_0$ and $k$ that best fit the observations (since we are assuming a constant core mass fraction γ, the only remaining variable in $\kappa$ is $k$). To find the best fit, we define the misfit function $\chi^2$:

$$\chi_\nu^2(x_0, k) = \frac{1}{\nu}\sum\left(\frac{f_{cc}-f_{cc,obs}}{\sigma}\right)^2 \quad (9)$$

where $f_{cc,obs}$ is the measured CC fraction for a particular element (Table 1), $\sigma$ is the associated uncertainty and $\nu$ is the number of degrees of freedom (4-2=2). We then perform a simple grid search to find the minimum value of $\chi_\nu^2$, i.e., the best fit. A value of $\chi_\nu^2 \approx 1$ indicates that the data are being fit at a level commensurate with the uncertainty; larger values indicate a poor fit and smaller values indicate over-fitting.

## 4. Results

### 4.1 Step Function

Figure 3 shows some example best-fit curves for the Earth with a pure NC main stage of accretion ($f_0$=0), a fixed, pure NC late veneer contribution ($x_1$=0.995, $f_2$=0) and $x_0$ held at different specified values: 0.5, 0.8, 0.9, 0.95, 0.96 (the best-fit value) and 0.98. Here the horizontal axis $x_{95}$ is a proxy for the partition coefficient $D$, where $x_{95}$ is defined as (Dauphas 2017)

$$x_{95} = (0.05)^{\frac{1}{\kappa+1}} \tag{10}$$

and represents the point in Earth's growth after which 95% of a particular element is recorded. A siderophile element (high $D$, high $\kappa$) means that $x_{95}$ is close to 1, so that the element records only the final stages of accretion. The curves decrease at high values of $x_{95}$ because we have assumed a pure-NC late veneer.

The $x_0$ values in Fig. 3 are chosen to illustrate the fundamental result: a low value of $x_0$ (e.g. 0.5 or 0.8) is obviously too small, while values in excess of 0.9 provide an improved fit to the data. Thus, we conclude that a CC-rich component must have been added to the Earth in the late stages of accretion.

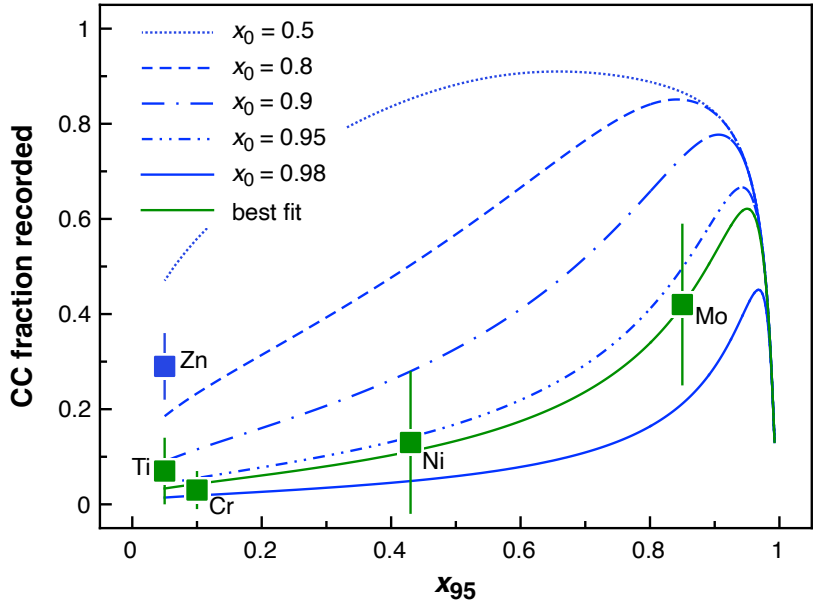

**Figure 3.** Green squares are CC fraction recorded by different non-volatile elements; blue square is volatile Zn (Table 1). Blue lines are best-fit curves for different specified values of $x_0$ using equation 5. The global best-fit value of $x_0$ is 0.96 (green line). Here $k$ is fixed at 0.2 (the value for the best-fit model), $f_0$ is fixed at 0 (pure NC), $f_1$ at 0.95 (CC-dominated) and the late veneer parameters $x_1$=0.995 and $f_2$=0 (pure NC). Note the poor fit for values of $x_0$ less than 0.9 ($\chi^2_\nu$ values are 119, 20.5, 4.44, 0.36, 0.20 and 1.27 for $x_0$=0.5, 0.8, 0.9, 0.95, 0.96 and 0.98).

We have so far assumed that the majority of material added to the planets was pure NC ($f_0 = 0$), as is the late veneer ($f_2$=0). However, Fig. 3 shows that the last ~10% of material added has to be more CC-rich. We therefore next explore how our results vary as we vary the NC fraction in the main stage of accretion ($f_0$) and the later stage ($f_1$) Fig. 4a shows how the misfit of the best-fit solution for Earth varies as $f_0$ and $f_1$ are varied. Misfits within error ($\chi^2_\nu < 1$) require $f_1$>0.3 and $f_0$<0.07. The best-fit value of $f_1$ is 0.95, so the later-added material could have been pure-CC, though the difference in misfit between $f_1$=0.3 and $f_1$=1 is small. Similarly, the best-fit value of $f_0$ is 0, so the main stages of Earth's accretion could have been from purely-NC materials. Figure 4b shows the corresponding best-fit value of $x_0$. For the regions with $\chi^2_\nu < 1$, above the dashed line, $x_0$ generally exceeds 0.9, while $k$ can range from 0.1 to 0.8 and is not well-constrained. The global best-fit value of $x_0$ is 0.96.

To summarize, assuming that the Earth's accretion can be represented as a simple three-step process (Fig. 2), the first 90-98% of material was NC-dominated (<7% CC) and potentially pure NC. Later-added material had to have a significant (>30%) CC component, while the final 0.5% of material added (the late veneer) is assumed to have been pure NC. The results do not change in any significant way if we had instead assumed that that late veneer was 80% NC.

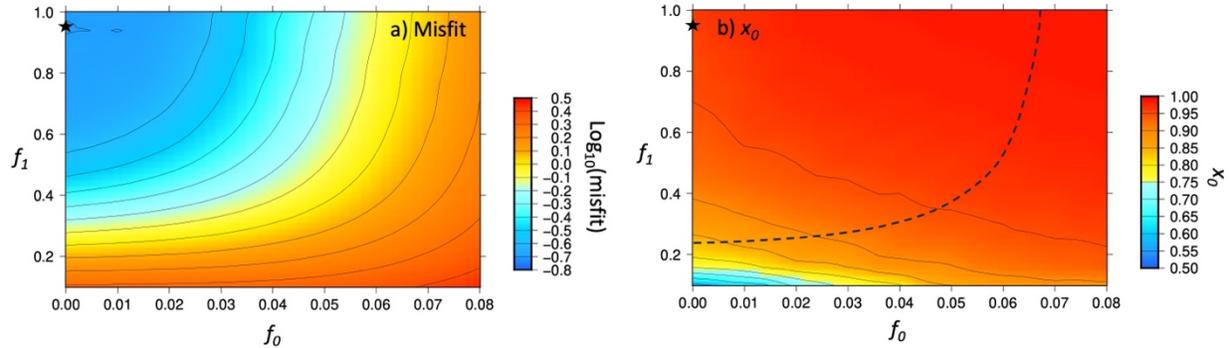

**Figure 4**. a) Variation in misfit (eq. 9) as a function of $f_0$ and $f_1$. Star denotes best-fit point ($f_0$=0, $f_1$=0.95, $k$=0.2, $x_0$=0.96, $\chi_\nu^2 = 0.20$). b) As for a) but showing the best-fit $x_0$ for each ($f_0,f_1$) combination. The dashed line shows the $\chi_\nu^2 = 1$ misfit line from a).

**4.2 Volatile elements**

So far we have assumed that only the CC fraction changes at $x_0$. A more general situation is the case when both the CC fraction *and* the concentration of elements delivered changes. Since the CC material is assumed to come from the outer solar system (Budde et al., 2016; Kruijer et al., 2017; Warren, 2011), it may well be the case that CC-rich material is also volatile-rich. For volatile elements, therefore, it is important to take any such changes in the concentration of elements delivered into account.

Figure 5 is similar to Fig. 3, but the blue line shows the effect of increasing the concentration of element delivered in the late-added material, where the ratio of the concentrations in late-delivered to early-delivered material is denoted $v_{fac}$. Since the late-delivered material is assumed to be CC-rich ($f_1$=0.95), the effect is to increase the CC fraction recorded for all elements. To fit the Zn results (blue square), we require $v_{fac} \approx 12$. Thus, the late-delivered CC-rich material is also volatile-enriched, by about an order of magnitude relative to the early-delivered material.

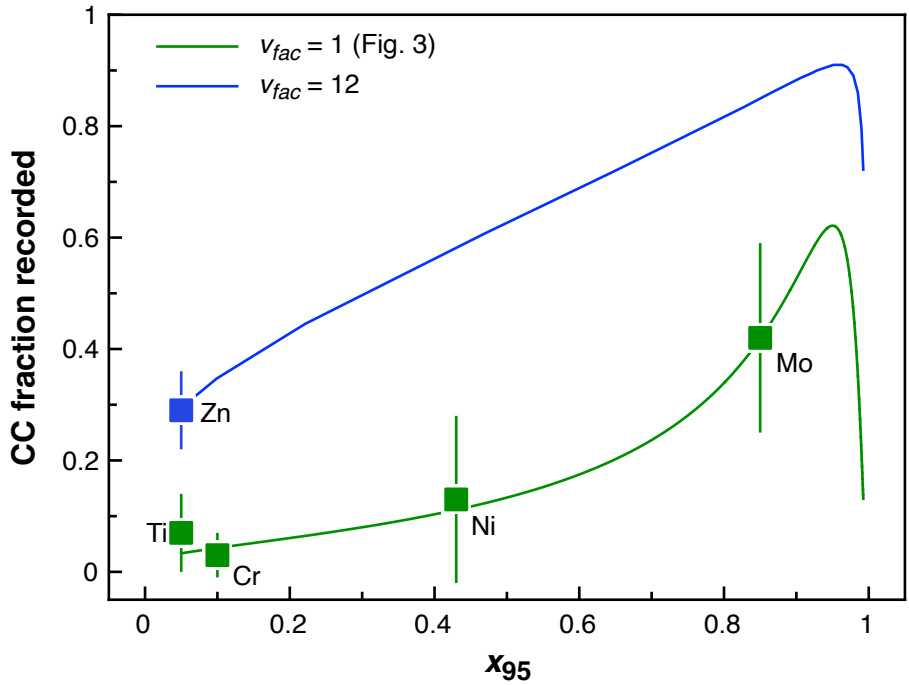

**Figure 5**. Green line is the best-fit model replotted from Figure 3. Blue line uses the same parameters, but with the CC-rich, late-delivered material (not including the final 0.5%) enhanced in concentration by a factor of 12. The Zn data (open blue square) suggests that this CC-rich material was volatile-rich.

The same result can be obtained directly from equation (6). For instance, if we take $f_0=0$, $f_1=0.95$ and $x_0=0.96$ (see Figure 3) then when $v_{fac}=11$, equation (6) gives a CC fraction of 0.30, which is the value recorded by Zn (Table 1). This result is also consistent with the relative concentrations in the NC and CC reservoirs deduced in constructing Figure 1 (38 and 309 µg/g, respectively).

## 5. Discussion

### 5.1 Comparison with prior studies

The key results of our study may be summarized as follows. The first 90-98% of material accreting to the Earth, as described by $x_0$, was NC-dominated ($f_0 < 0.07$), while the later material added was markedly enriched in CC ($f_1 > 0.3$) and could have been an NC/CC mix or pure CC. This late-delivered CC-rich material was also enriched in volatile elements by roughly an order of magnitude ($v_{fac} \approx 10$) relative to the early-delivered material. Finally, despite the late delivery of CC material, the late veneer was again dominated by NC material. Several of these observations have been made in prior studies, but usually these studies were focused on addressing one specific aspect of Earth's accretion history, such as for instance the bulk CC fraction in Earth (e.g., Burkhardt et al., 2021; Schiller et al., 2018; Warren, 2011) or the origin of the late veneer (e.g., Bermingham et al., 2018; Fischer-Gödde and Kleine, 2017; Worsham and Kleine, 2021). The present study is most similar to that of Dauphas et al. (2024), which also investigated how the provenance of Earth's building material evolved over time. Despite differences in methodology, several of our findings are consistent with those of Dauphas et al. (2024), although there are also some notable differences, as follows.

Both studies find an Earth consisting overall of about 92-94% NC-like material and only 6-8% CC-like material, and also find that the fraction of CC material delivered to Earth increased with increasing time. However, while Dauphas et al. (2024) argue that the CC material in Earth was accreted during the last 40% of its growth, we define this time interval much more precisely to <10%. This difference reflects that Dauphas et al. (2024) fix the boundaries between different accretion stages *a priori*, while we allow them to vary. Evidently, this makes it possible to determine the time of CC delivery to Earth more precisely, which in turn has important implications for understanding the process by which Earth accreted this material (see below). In particular, it becomes important to consider the relationship between the CC-rich material and the Moon-forming impactor, which we do in Section 5.3 below. Another important difference is that Dauphas et al. (2024) remain agnostic on the details of Zn delivery to the Earth while we conclude that the late-added CC-rich material was enriched in volatiles such as Zn by a factor of ~10 relative to the early-delivered material.

**5.2 Nature and timing of CC delivery to the inner solar system**

The key result of our model is that the late addition to the Earth of a few percent of pure-CC material can match the available data. This conclusion also holds true for volatile Zn if the late-delivered CC-rich material was also volatile-enriched (Figure 5). Of note, the late-stage addition of volatile-enriched material is consistent with other lines of evidence, notably Pd-Ag isotope systematics (Schönbächler et al., 2010) and coupled accretion-differentiation models (Rubie et al., 2015). The inward scattering of CC bodies is thought to be related to the growth and migration of the gas giant planets, which must have happened within the lifetime of the protoplanetary disk, i.e., within less than ~5 million years after solar system formation (Raymond and Izidoro, 2017). However, accretion of the Earth took much longer, several tens of millions of years, and thus extended well beyond the lifetime of the disk (e.g., Nimmo and Kleine, 2015). As such, our finding that CC material was accreted by Earth predominantly or even exclusively during the last few percent of Earth's growth appears inconsistent with the expectation that CC bodies were scattered into the inner solar system early.

There are two observations that help understanding this apparent conundrum. First, Mars appears to have accreted a much lower fraction of CC material than Earth did (Kleine et al., 2023; Paquet et al., 2023). This is surprising, because if the CC material was accreted predominantly by small bodies (planetesimals), then dynamical models predict that Mars should have accreted a similar or even larger share of CC objects than Earth (Raymond and Izidoro, 2017; Walsh et al., 2011). Second, the Earth's late veneer seems to predominantly consist of NC bodies with no detectable CC contribution (<20% CC, see above), indicating an overall very low fraction of CC planetesimals in the terrestrial planet region, which in turn is consistent with the low CC fraction found in Mars. Combined, these two observations indicate that the fraction of CC planetesimals scattered into the inner solar system during the growth and migration of the giant planets was low and left no detectable signature either in the isotopic signature of Mars or of the terrestrial late veneer. Consequently, the larger CC fraction in Earth cannot reflect the accretion of a background swarm of small CC planetesimals, but could instead result from the stochastic accretion of few larger CC objects (embryos) that were not accreted by Mars and also did not contribute significantly to the late veneer.

Any successful model for the accretion of Earth must account for the observation that Earth accreted about 6% of its total mass from CC material, of which much was acquired late in its growth history, and that this material was added by large bodies and not by planetesimals. Dauphas et al. (2024) recently proposed the idea that a single or a few CC interlopers of approximately lunar mass hit the Earth during the course of its growth and are the main sources of the CC material in Earth. We prefer the scenario where multiple CC embryos were involved because in this case, statistically, it is plausible that one of these objects hit the Earth relatively late, consistent with the observation that the CC addition to Earth occurred late. Because the total amount of CC mass accreted by Earth was only ~6%, the last of these CC impactors could not have brought more than a few percent of the Earth's mass. Importantly, this is consistent with our $x_0$ values derived above. In order to deliver a sufficient fraction of the BSE's Mo, this object either was undifferentiated, the core equilibrated efficiently with the Earth's mantle, or most of its core remained stranded in the terrestrial mantle (Korenaga and Marchi, 2023).

As an example, we consider here undifferentiated CC impactors (identical to assuming complete impactor core re-equilibration), each around 0.02 $M_E$. While an exhaustive sweep of parameter space is beyond the scope of this study, we show in Fig. 6 that adding three of these pure-CC bodies at regular intervals (e.g., when Earth is at 32%, 65% and 97.5% of its growth) to an otherwise pure-NC Earth is consistent with the observations. Note that this scenario assumes a pure NC composition of the late veneer and that there is no enhancement in CC concentrations (i.e. $v_{fac}$=1). This plot also shows what happens if the final 0.02 $M_E$-sized body is not included (dashed line), demonstrating that a late, CC-rich impactor is essential.

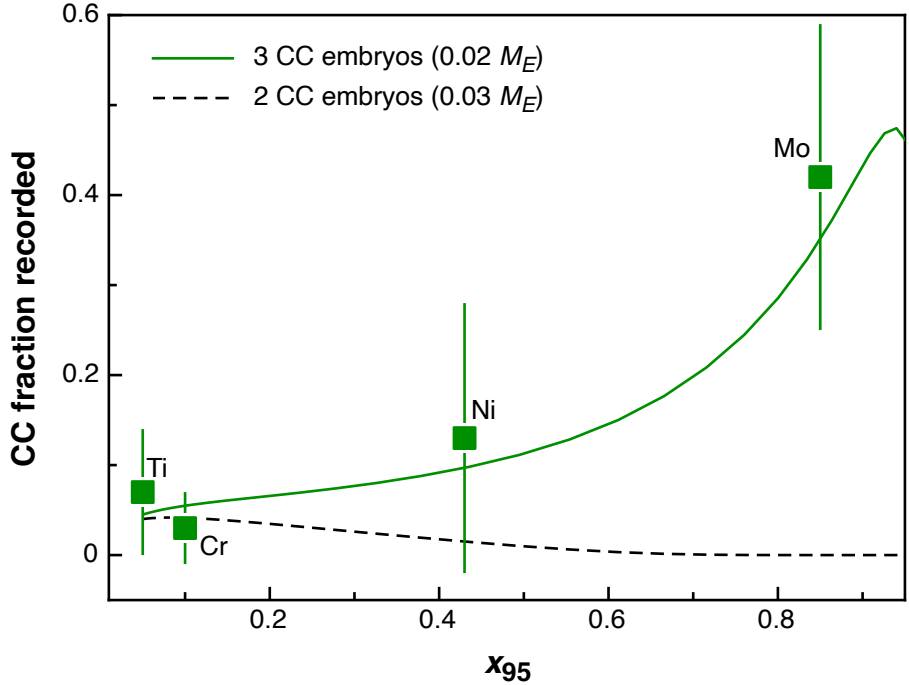

**Figure 6.** Green line shows CC fraction as a function of $x_{95}$ due to the addition of three 0.02 $M_E$ bodies, calculated using equation (8). These are pure CC ($f_1$=1) and are added at 0.32, 0.65 and 0.975 of the Earth's growth; the other material added (including the last 0.5% of accretion, i.e. the late veneer) is pure NC ($f_0$=0). Here $k$=0.2. Black dashed line shows the effect of adding only the first two bodies, now increased in mass to 0.03 $M_E$ each. Green squares are the non-volatile elements re-plotted from Figure 3 (see Table 1).

In summary, our results are consistent with the idea that a handful of CC, roughly Moon-mass bodies were injected into the terrestrial planet-forming region, and suggests that there was only a negligible total mass of smaller CC planetesimals in that region. The latter accounts for the low CC fraction in Mars and in the late veneer. This distribution of CC objects in the inner solar system may either reveal a top-heavy mass distribution of objects formed in the giant planet region, or be a consequence of gas drag, during which the orbits of small bodies (planetesimals) become rapidly circularized, leading to a larger mass of CC planetesimals implanted in the asteroid belt than reaching the terrestrial planet region (Raymond and Izidoro, 2017). Moon-size bodies, instead, are not very sensitive to gas drag, so they can "overshoot" the asteroid belt and land in the terrestrial planet region.

**5.3 Role of the Moon-forming impact**

The scenario envisaged for Fig. 6, although *ad hoc*, satisfies the constraints imposed by (*i*) the inferred CC fractions recorded by different elements and the associated late CC delivery to Earth, (*ii*) the low CC fraction in Mars, and (*iii*) the NC-rich nature of the late veneer. However, modeling the late stages of Earth growth must also consider the Moon-forming impactor (Theia), which is commonly thought to have been the last large body that struck the Earth (Canup and Asphaug, 2001). As shown by Budde et al. (2019), the isotopic heritage of Theia has a profound effect on the BSE's isotope composition of strongly siderophile elements such as Mo. This is because the BSE's Mo records only the last ~10% of Earth's accretion, which corresponds to the size of Theia assumed in the canonical model of the giant impact (Canup and Asphaug, 2001).

There are four scenarios that can be envisioned for how the late-stage CC delivery to Earth is linked to or is influenced by the formation of the Moon. The first is that Theia itself is the last CC embryo to hit the Earth. Budde et al. (2019) have shown that the BSE's mixed NC-CC Mo isotope signature can be accounted for by invoking a CC origin for Theia, especially if a pure NC composition is assumed for the late veneer (see their Fig. 4). However, these authors also showed that the overall effect of the Moon-forming impact on the BSE's Mo isotope composition strongly depends on the degree to which Theia's core equilibrated with proto-Earth's mantle. This question can be assessed based on fluid dynamics experiments; inspection of Fig. 10 of Landeau et al. (2021) shows that siderophile elements with $D$~100 (appropriate for Mo) will not be subject to any significant equilibration during the passage of the impactor core. In this case, the results of Budde et al. (2019) suggest that the Moon-forming impactor was an NC and not a CC body.

The second possibility is that the Moon-forming impactor was NC and happened before the last CC embryo hit the Earth. This scenario would require efficient equilibration of that embryo's core with Earth's mantle, because otherwise the subsequent siderophile CC signature in the BSE would be too small. Thus, as before the feasibility of this scenario depends on our understanding of impactor core re-equilibration during large impacts. Going back to Fig. 10 of Landeau et al.

(2021) shows that for impact of a ~0.02 $M_E$ embryo little equilibration is expected, so we view this scenario as unlikely.

The third possibility is that the Moon-forming impactor was NC and happened after the last CC embryo. As noted above, the CC fraction recorded in the siderophile elements could be retained through such a Moon-forming impact, as long as the degree of impactor core re-equilibration is low (Budde et al., 2019). For instance, if Theia's core only equilibrated with one-third the target mantle, that third would acquire the NC isotopic signature of the siderophile elements contained in Theia's core, while the other two-thirds would retain the previous siderophile CC signature. Subsequent mixing of these two reservoirs would then yield the observed mixed NC-CC signature of the BSE's Mo. This scenario is consistent with the expectation that the effective equilibration factor will be smaller for more siderophile elements (see Section 2), and will be smaller for larger impactors (see above and Landeau et al., 2021). In this picture, the partial NC signature for Mo favours limited equilibration and thus a larger Moon-forming impactor.

Finally, a fourth possibility is that the Moon-forming impactor, although being a primarily NC body, had a CC-rich signature in its mantle, acquired from a prior collision with an CC embryo. Since the degree of equilibration between Theia's core and Earth's mantle is expected to be low (see above), some of this CC-rich signature would be imparted into the isotopic composition of the BSE.

In all the four scenarios, Earth's pre-late veneer mantle (i.e., the mantle after the Moon-forming impact but before late accretion) had a mixed NC-CC signature. Fischer-Gödde et al. (2020), however, argued that the pre-late veneer mantle was purely NC. These authors showed that samples from Isua are characterized by an *s*-process-enriched Ru isotope composition relative to the BSE. As *s*-process-enriched material is commonly associated with the innermost disk (Burkhardt et al., 2021), this composition indicates a pure NC origin of the Ru in these rocks. If this composition were indeed a pre-late veneer signature, it would imply that the BSE's pre-late veneer Mo isotope signature also was purely NC, and not mixed NC-CC as we inferred above. However, while the Ru isotope data suggest that the source of the Isua rocks did not receive the full complement of the late veneer, it may have well received parts of the late veneer. In this case, the Ru isotope anomalies would not record a pre-late veneer signature, but would instead indicate that the late veneer itself consisted of objects with varying isotopic compositions and is likely NC-dominated. Thus, we regard a late impact of a CC embryo followed by a final, Moon-forming impact of a primarily NC body (which may itself also contain some CC material) and a NC-dominated late veneer as the most probable scenario.

**5.4 Model for CC accretion to the terrestrial planets**

Figure 7 sketches our preferred scenario for the delivery of CC material into the inner disk, and illustrates the envisaged sequence of processes by which this material was accreted to the terrestrial planets. The terrestrial planets start accreting from pure NC material, as indicated by the overall low CC fraction in Earth and Mars, and the observation that for most elements Earth is isotopically very similar to enstatite chondrites (Fig. 7a). The early growth and migration of the giant planets leads to scattering of CC bodies into the inner disk, with most of the mass of these bodies in a few large CC embryos. Due to gas drag, the CC objects implanted into the asteroid belt are predominantly planetesimals, while the larger embryos are injected into the terrestrial planet region (Fig. 7b). This process accounts for the overall low CC fraction in Mars and the terrestrial late veneer, indicating a generally low fraction of CC planetesimals in the terrestrial planet region. Earth accretes a few of the CC embryos, including one late in the accretionary sequence, while Mars, owing to its much smaller size, did not (Fig. 7c). This accounts for the overall larger CC fraction in Earth compared to Mars and the increased CC addition late in Earth's growth. Finally, the Earth is struck by one final large NC-dominated object leading to the formation of the Moon, after which it acquires ~0.5% of NC-dominated material as the late veneer (Fig. 7d).

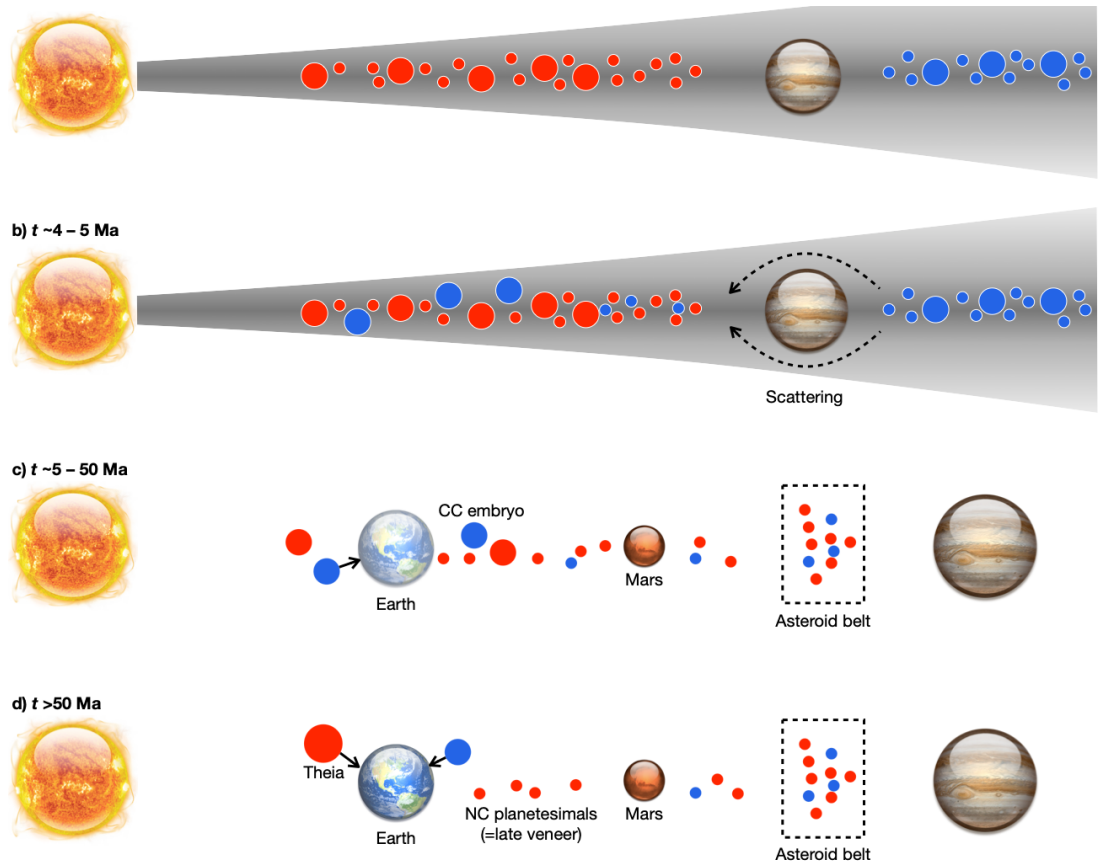

**Fig. 7.** Cartoon illustrating how the provenance of Earth's building material evolved over time with a particular emphasis on how and when CC material was delivered to the terrestrial planets. a) Initial accretion in the terrestrial planets is of pure-NC material. b) Early giant planet growth/migration scatters a few large CC embryos into the inner disk while CC planetesimals are implanted into the asteroid belt. c) Earth accretes a few CC embryos as it grows, including one late in the accretionary sequence, but Mars does not. d) The Earth suffers one final NC-dominated, Moon-forming impact, after which it acquires ~0.5% of NC-dominated material (the late veneer). Timing given for each panel is not strict and mostly for illustrative purposes.

## 6. Conclusions

This study reconstructs the accretion history of the Earth by utilizing a comprehensive set of nucleosynthetic isotope anomaly data for meteorites and terrestrial samples reported in numerous prior studies. Comparing the nucleosynthetic isotope signatures of the BSE and Mars to those of NC and CC meteorites provides the following three key observations: (*i*) Earth accreted predominantly from NC material with only small contributions (6±4%) of CC material, much of which was delivered late, during the last 2-10% of Earth's accretion and which was a factor of ~10 richer in moderately volatile elements compared to the early-accreted material; (*ii*) despite this late delivery of CC material to Earth, the last ~0.5% of Earth's accretion (i.e., late veneer) was NC-dominated with no detectable signature of CC material; and (*iii*) Mars contains much less CC material than Earth.

These three observations combined can be reconciled if the CC bodies that were delivered to the inner solar system as the giant planets grew had the bulk of their mass concentrated in a few, Moon-sized embryos rather than in smaller planetesimals. This readily accounts for the low fraction of CC material in both Mars and the terrestrial late veneer, which predominantly accreted CC material through planetesimals. Earth, by contrast, received a larger share of CC material through accretion of a few large CC objects for which, owing to their large size, the collision probability with Mars was very low. This scenario is consistent with the expected effects of gas drag on the orbits of inwards scattering objects, which led to the preferential implantation of planetesimals into the asteroid belt and of embryos into the terrestrial planet region.

Because CC bodies were likely injected in the terrestrial planet region early, it is statistically highly unlikely that Earth was hit by CC bodies exclusively at the end of its growth. Fig. 6 shows that addition of a few CC bodies during Earth's growth, with one happening late, is consistent with the observations. It is also unlikely that the CC material was supplied by the Moon-forming impact itself, because the degree of impactor core equilibration was likely small. Instead, the most probable scenario appears to be that the last CC impact occurred prior to the Moon-forming impact of a large NC body, as long as the latter did not re-homogenize the entire Earth's mantle.

Although the scenario presented in Fig. 7 is consistent with the isotopic observations, it has not yet been demonstrated to work dynamically. That is beyond the scope of this paper, but is an obvious next step. The advantage of such dynamical simulations is that not only do they track the provenance of accreting objects, but they can also track when each impact occurs, which can then be used to predict the final CC fraction recorded for different elements. As such, combining dynamical simulations with the isotopic constraints used here would allow us to assess whether CC embryos are indeed preferentially injected into the terrestrial planet region, and whether Earth accreted at least one of these objects while Mars did not accrete any of them, as we would predict.

## Acknowledgements

FN acknowledges support from NSF-CSEDI-2054876. TK and AM acknowledge support from the ERC HolyEarth project. We thank Sean Raymond, Rich Walker and one anonymous reviewer for careful reading of this manuscript.